\documentclass[aps,prd,onecolumn,preprintnumbers,groupedaddress,showpacs,nofootinbib,amssymb]{revtex4}
\usepackage{graphicx}
\usepackage{amsmath}
\usepackage{amssymb}
\usepackage{amsfonts}
\usepackage{bm}

\begin{document}
 
\makeatletter
\renewcommand{\theequation}{\Roman{section}.\arabic{equation}}
\@addtoreset{equation}{section}
\makeatother

\def\pp{{\, \mid \hskip -1.5mm =}}
\def\cL{\mathcal{L}}
\def\be{\begin{equation}}
\def\ee{\end{equation}}
\def\bea{\begin{eqnarray}}
\def\eea{\end{eqnarray}}
\def\tr{\mathrm{tr}\, }
\def\nn{\nonumber \\}
\def\e{\mathrm{e}}

\newcommand{\Eqn}[1]{&\hspace{-0.2em}#1\hspace{-0.2em}&}
\newcommand{\abs}[1]{\vert{#1}\vert}
\def\Vec#1{\mbox{\boldmath $#1$}}
\def\Vecs#1{\mbox{\boldmath\tiny $#1$}}
\def\Lap{{\mathop{\Delta}\limits^{(3)}}}

\title{Trace-anomaly driven inflation in $f(T)$ gravity and in minimal massive bigravity}
\author{Kazuharu Bamba$^{1,}$,
Shin'ichi Nojiri$^{1, 2,}$ and 
Sergei D. Odintsov$^{2, 3, 4, 5}$}
\affiliation{
$^1$Kobayashi-Maskawa Institute for the Origin of Particles and the
Universe,
Nagoya University, Nagoya 464-8602, Japan\\
$^2$Department of Physics, Nagoya University, Nagoya 464-8602, Japan\\
$^3$Instituci\`{o} Catalana de Recerca i Estudis Avan\c{c}ats (ICREA),
Barcelona, Spain\\
$^4$Institut de Ciencies de l'Espai (CSIC-IEEC),
Campus UAB, Facultat de Ciencies, Torre C5-Par-2a pl, E-08193 Bellaterra
(Barcelona), Spain\\
$^5$ Tomsk State Pedagogical University, Kievskaya Avenue, 60,
634061, Tomsk, Russia}


\begin{abstract} 
We explore trace-anomaly driven inflation in modified gravity. 
It is explicitly shown that in $T^2$ teleparallel gravity, 
the de Sitter inflation can occur, although quasi de Sitter inflation happens in $R^2$ gravity. Furthermore, we investigate the influence of the trace anomaly on inflation. It is found that in $f(T)$ gravity, the de Sitter inflation can end because it becomes unstable due to the trace anomaly, whereas also in higher derivative gravity, the de Sitter inflation can be realized and it will be over thanks to the trace anomaly for smaller parameter regions in comparison with those in teleparallelism. The instability of the de Sitter inflation in $T^2$ gravity and $R^2$ gravity (both with taking account of the trace anomaly) is examined. In addition, we study trace-anomaly driven inflation in minimal massive bigravity, where the contribution from the massive graviton acts as negative cosmological constant. It is demonstrated that the de Sitter inflation can occur and continue for long enough duration. 
\end{abstract}


\maketitle

\section{Introduction}

For a long time, it is expected that vacuum quantum effects and/or 
quantum gravity effects should play an important role for the early-time cosmological evolution~\cite{1,2}. 
It was realized in~\cite{D} that quantum effects due to conformally invariant 
fields (conformal or trace-anomaly induced effective action) which is expected 
to be most relevant at the early universe may lead to the early-time de Sitter 
universe. This may be considered to be an example of an eternal inflation model. Nevertheless, making more deep analysis and adjusting the coefficient of 
higher (total) derivative counter-term in the trace-anomaly, 
the very successful model of trace-anomaly driven inflation (also 
called Starobinsky inflation~\cite{star}) has been 
constructed (for its extended discussion, see~\cite{alex}). 
The $\Box R$ term in the trace anomaly permits the long-lived inflation with the possibility of the exit~\cite{star}. Note that in the anomaly-induced action such a term is a higher-derivative $R^2$ term.
Note also that the trace-anomaly induced action is just an effective gravitational action with $R^2$-term and non-local term, so it may be considered as natural contribution to modified gravity. 
Furthermore, trace-anomaly driven inflation has been extended for gravity with 
anti-symmetric torsion~\cite{buc}, dilaton-coupled gravity~\cite{no} where new 
dilaton dependent terms appear in the conformal anomaly and brane-world 
gravity~\cite{hawking, hawking1} 
where it was called Brane New World.

Recent cosmological data~\cite{WMAP, 
Hinshaw:2012aka, Ade:2013lta, Ade:2013uln} 
indicate that trace-anomaly driven inflation or its closely related $R^2$-version may be one of the most realistic candidates of inflationary theory. 
This has attracted a number of researchers to study different aspects of 
$R^2$ inflation (for a very incomplete list of related papers, 
see~\cite{star1-SCMOZ-S-Inf} and references therein). 

In this Letter, we investigate trace-anomaly driven inflation 
in modified gravity of teleparallel type~\cite{A-P} and in massive bigravity. 
In particular, we study the analog of an $R^2$ inflation 
model~\cite{star, star1-SCMOZ-S-Inf} 
in extended teleparallelism, namely, so-called $f(T)$ gravity. 
The main reason why $f(T)$ gravity has vigorously been 
explored in the literature is that 
current cosmic acceleration~\cite{Bengochea:2008gz, Linder:2010py, WY-BGLL-BGL, Geng:2011aj} as well as inflation~\cite{Inflation-F-F} 
can occur in $f(T)$ gravity. Hence, various accelerating cosmology features of 
$f(T)$ gravity itself have also been explored, e.g., in Refs.~\cite{CP-FT, Li:2010cg, TG-S, OINC-IGO, Yang:2010ji, Bamba:2013jqa, Basilakos:2013rua, DHA-BDHO, M-BGL}. 
We find that for $T^2$ gravity, there can exist the de Sitter solution. 
This is different from the fact that in usual $R^2$ gravity 
a quasi de Sitter inflation can happen. 
Furthermore, we examine the influence of the trace anomaly 
on $T^2$ inflation and demonstrate that since the de Sitter solution is not 
stable, inflation can be over. In other words, by taking the trace anomaly 
into consideration, the so-called graceful exit 
problem from inflation can be solved as in seminal paper~\cite{star}. 
We also observe that in convenient higher derivative gravity, 
the de Sitter inflation can be realized and finally it can end 
owing to the existence of the trace anomaly, although 
for much smaller parameter regions than those in teleparallelism.  
Moreover, we explore trace-anomaly driven inflation in other modified gravity, 
i.e., in minimal massive bigravity. 
It has recently been presented the non-linear ghost-free massive 
gravity (de Rham, Gabadadze, and Tolley (dRGT)) model~\cite{DRG-DRGT, Hassan:2011hr}). The dRGT model has non-dynamical background, but models with dynamical 
metric have been found in Ref.~\cite{HR-HR} (for a recent very detailed review of massive gravity, see~\cite{deRham:2014zqa}). 
It is shown that the de Sitter inflation with long enough duration 
can occur, and that its instability can lead to the end of the de Sitter 
inflationary stage. 
We use units of $k_\mathrm{B} = c = \hbar = 1$, 
where $c$ is the speed of light, and denote the
gravitational constant $8 \pi G_{\mathrm{N}}$ by
${\kappa}^2 \equiv 8\pi/{M_{\mathrm{Pl}}}^2$
with the Planck mass of $M_{\mathrm{Pl}} = G_{\mathrm{N}}^{-1/2} = 1.2 \times
10^{19}$\,\,GeV. 

The Letter is organized as follows.
In Sec.\ II, we briefly describe the formulation of teleparallelism. 
In Sec.\ III, we explore (quasi-de Sitter) inflation in $R^2$ gravity in the Einstein frame. The comparison with $T^2$ gravity is made, where only eternal de Sitter inflation is possible. 
Also, in Sec.\ IV, we investigate the trace-anomaly driven inflation in $f(T)$ gravity. We show that in $f(T)$ gravity, taking account of the trace anomaly 
makes the exit possible like in usual $R^2$ gravity. 
Furthermore, we study trace-anomaly driven inflation in minimal massive bigravity in Sec.\ V, where it is demonstrated that the massive term acts as negative cosmological constant. 
Finally, conclusions are given in Sec.\ VI. 

\section{Teleparallelism}

We present the essence of teleparallelism~\cite{A-P}. 
We introduce orthonormal tetrad components, i.e., 
the so-called viervein fields $e_a (x^{\mu})$, where $a = 0, \cdots, 3$ 
with the roman index $a$ denoting the tangent space of 
a manifold at a point $x^{\mu}$. 
We describe the metric with the tetrad components as 
$g_{\mu\nu}=\eta_{a b} e^a_\mu e^b_\nu$, 
where the Greek indices $\mu, \, \nu$ $(= 0, \cdots, 3)$ 
show coordinates on the manifold, and 
hence $e_a^\mu$ can be regarded as its tangent vector. 
Moreover, the inverse of the viervein fields can be 
derived from the equation 
$e^a_\mu e_a^\nu = \delta_\mu^\nu$. 
By using the Weitzenb\"{o}ck connection 
$\Gamma^{(\mathrm{W})\, \rho}_{\verb|   |\nu\mu} \equiv e^\rho_a \partial_\mu e^a_\nu$, we define the torsion tensor as 
$T^\rho_{\verb| |\mu\nu} \equiv 
\Gamma^{(\mathrm{W})\,\rho}_{\verb|   |\nu\mu} - \Gamma^{(\mathrm{W})\,\rho}_{\verb|   |\mu\nu} = e^\rho_a 
\left( \partial_\mu e^a_\nu - \partial_\nu e^a_\mu \right)$. 
In this case, curvature vanishes and only non-zero torsion exists. 

To construct the torsion scalar, we further make the superpotential 
$S_\rho^{\verb| |\mu\nu}$ and the contorsion tensor $K^{\mu\nu}_{\verb|  |\rho} $, given by 
\begin{equation} 
S_\rho^{\verb| |\mu\nu} \equiv 
\frac{1}{2}
\left(K^{\mu\nu}_{\verb|  |\rho}+\delta^\mu_\rho \ 
T^{\alpha \nu}_{\verb|  |\alpha}-\delta^\nu_\rho \ 
T^{\alpha \mu}_{\verb|  |\alpha}\right)\,,
\quad 
K^{\mu\nu}_{\verb|  |\rho} 
\equiv 
-\frac{1}{2} 
\left(T^{\mu\nu}_{\verb|  |\rho} - T^{\nu \mu}_{\verb|  |\rho} - 
T_\rho^{\verb| |\mu\nu}\right)\,.
\label{eq:add-II.01}
\end{equation}
Eventually, we can build the torsion scalar as 
\begin{equation} 
T \equiv S_\rho^{\verb| |\mu\nu} T^\rho_{\verb| |\mu\nu} = 
\frac{1}{4}
\left(T^{\rho\mu\nu}T_{\rho\mu\nu} + 
2T^{\rho\mu\nu}T_{\nu\mu\rho}-4T_{\rho\mu}^{\verb|  |\rho}
T^{\nu\mu}_{\verb|  |\nu}\right)\,. 
\label{eq:add-II.02}
\end{equation}

The action of extended teleparallelism, namely, $f(T)$ gravity, 
is written as 
\begin{equation} 
S= \int d^4 x \abs{e} \frac{f(T)}{2{\kappa}^2} 
+ S_{\mathrm{matter}} \,, 
\label{eq:4.1}
\end{equation}
with the matter action 
\begin{equation} 
S_{\mathrm{matter}} = 
\int d^4 x {\mathcal{L}}_{\mathrm{matter}} 
\left( g_{\mu\nu}, {\Phi}_i \right)\,. 
\label{eq:4.2}
\end{equation}
Here, we express the determinant of the metric $g_{\mu \nu}$ 
as $\abs{e}= \det \left(e^a_\mu \right)=\sqrt{-g}$, 
and ${\mathcal{L}}_{\mathrm{matter}} \left( g_{\mu\nu}, {\Phi}_i \right)$ with 
${\Phi}_i$ matter fields is the Lagrangian of matters. 
If $f(T) = T$, this theory is pure teleparallelism. 

It follows from the action in Eq.~(\ref{eq:4.1}) 
that we have~\cite{Bengochea:2008gz} 
\begin{equation}
\frac{1}{e} \partial_\mu \left( eS_a^{\verb| |\mu\nu} \right) f^{\prime} 
-e_a^\lambda T^\rho_{\verb| |\mu \lambda} S_\rho^{\verb| |\nu\mu} 
f^{\prime} +S_a^{\verb| |\mu\nu} \partial_\mu \left(T\right) f^{\prime\prime} 
+\frac{1}{4} e_a^\nu f = \frac{{\kappa}^2}{2} e_a^\rho 
T_{\mathrm{matter}\, \rho}^{\verb|      |\nu}\,, 
\label{eq:2.7}
\end{equation}
with $T_{\mathrm{matter}\, \rho}^{\verb|      |\nu}$ 
the energy-momentum tensor of matters 
and the prime denoting the derivative with respect to $T$. 
Also, the relation of $R$ to $T$ is given by~\cite{A-P} 
\begin{equation}
R = -\left( T + 2\nabla^{\mu} T^{\rho}_{\verb| |\mu \rho} \right)\,,  
\label{eq:ADD-FT-trace-1-a-II.5}
\end{equation}
where $\nabla^{\mu}$ is the covariant derivative in terms of the metric tensor 
$g_{\mu\nu}$. 
It shows that General Relativity is equivalent to $T$-gravity, while $f(T)$ 
gravity is not a covariant theory because only a piece of curvature is kept in 
$T$. 

\section{$R^2$ inflationary model} 

\subsection{$R^2$ gravity} 

The action of an $R^2$ theory is given by\footnote{For the action 
representing an $R^2$ term as $S = \int d^4 x \sqrt{-g} R^2$, 
the trace of the energy-momentum tensor is defined by 
$g^{\mu\nu} \left(2/\sqrt{-g} \right) 
\left(\delta S/\delta g^{\mu\nu}\right)$, 
and this gives $\Box R$. 
Then, effectively the $R^2$ term in 
Eq.~(\ref{eq:ADD-FT-trace-1-a-III.1}) 
comes from $\Box R$ in the trace anomaly. 
Furthermore, the coefficient of the $\Box R$ term is not fixed 
in the trace anomaly and the trace anomaly produces one more non-local term in the effective gravitational action. Nevertheless, thanks to the above relation between $R^2$ and $\Box R$, $R^2$ inflation is qualitatively very similar to trace-anomaly driven inflation.}
\begin{equation}
S = \int d^4 x \sqrt{-g} \frac{1}{2\kappa^2}
\left(R+\frac{1}{6M_{\mathrm{S}}^2} R^2 \right)\,,  
\label{eq:ADD-FT-trace-1-a-III.1}
\end{equation}
where $M_{\mathrm{S}}$ is a mass scale. 
Substituting the relation in Eq.~(\ref{eq:ADD-FT-trace-1-a-II.5}) 
into the action in Eq.~(\ref{eq:ADD-FT-trace-1-a-III.1}), we find 
%
\begin{eqnarray} 
S \Eqn{=} \int d^4 x \sqrt{-g} \frac{1}{2\kappa^2} 
\left[ -\left( T + 2\nabla^{\mu} T^{\rho}_{\verb| |\mu \rho} \right) 
+ \frac{1}{6M_{\mathrm{S}}^2} 
\left( T + 2\nabla^{\mu} T^{\rho}_{\verb| |\mu \rho} \right)^2 
\right]  \nonumber \\ 
\Eqn{=} 
\int d^4 x \sqrt{-g} \frac{1}{2\kappa^2} 
\left\{ -T + \frac{1}{6M_{\mathrm{S}}^2} T^2 
+\frac{2}{3M_{\mathrm{S}}^2} 
\left[T \nabla^{\mu} T^{\rho}_{\verb| |\mu \rho}
+\left( \nabla^{\mu} T^{\rho}_{\verb| |\mu \rho} 
\right)^2 \right] \right\}\,,
\label{eq:ADD-FT-trace-1-a13-IIIA-001}
\end{eqnarray} 
%
where in deriving Eq.~(\ref{eq:ADD-FT-trace-1-a13-IIIA-001}), 
we have done a partial integration. 
By comparing the action in Eq.~(\ref{eq:ADD-FT-trace-1-a-III.1}) 
with that in Eq.~(\ref{eq:ADD-FT-trace-1-a13-IIIA-001}), 
we see the difference of the action of $R^2$ gravity from that of 
$T^2$ gravity. 

In general, the action of $F(R)$ gravity is represented as 
%
\begin{equation}
S = \int d^4 x \sqrt{-g} \frac{F(R)}{2\kappa^2} \,.
\label{eq:ADD-FT-trace-1-a-III-B-01}
\end{equation}
For the action in Eq.~(\ref{eq:ADD-FT-trace-1-a-III.1}), we have 
$F(R) = R+\left[ 1/\left(6M_{\mathrm{S}}^2 \right) \right] R^2$. 
With an auxiliary field $\psi$, 
the action in Eq.~(\ref{eq:ADD-FT-trace-1-a-III-B-01}) is rewritten to 
\begin{equation} 
S = \int d^4 x \sqrt{-g} \frac{1}{2\kappa^2} 
\left[
F(\psi) + \left(R-\psi \right) \frac{d F(\psi)}{d \psi} \right]\,.
\label{eq:FR9-2-IVB2-01} 
\end{equation} 
By varying this action with respect to $\psi$, we find 
$\left(R-\psi \right) d^2 F(\psi)/d \psi^2 = 0$. 
Provided that $d^2 F(\psi)/d \psi^2 \neq 0$, we obtain $R = \psi$. 
Combining this relation with the action in Eq.~(\ref{eq:FR9-2-IVB2-01}) 
leads to the original action of $F(R)$ gravity with matters 
in Eq.~(\ref{eq:ADD-FT-trace-1-a-III-B-01}). 
Moreover, by defining $\vartheta \equiv d F(\psi)/d \psi$, we 
express the action in Eq.~(\ref{eq:FR9-2-IVB2-01}) as 
\begin{eqnarray} 
S \Eqn{=} \int d^4 x \sqrt{-g} \left( \frac{\vartheta R}{2\kappa^2} 
-\Upsilon (\vartheta) \right) \,,
\label{eq:FR9-2-IVB2-02} \\
\Upsilon (\vartheta) \Eqn{\equiv} \frac{1}{2\kappa^2} 
\left( \vartheta (\psi) \psi - F(\vartheta (\psi)) \right)\,.
\label{eq:FR9-2-IVB2-03}
\end{eqnarray} 
Through the conformal transformation 
$g_{\mu\nu} \to \hat{g}_{\mu\nu} \equiv \Omega^2 g_{\mu\nu}$ with 
$\Omega^2 = \vartheta$, we acquire~\cite{CT-F-M}
\begin{eqnarray} 
S \Eqn{=} \int d^4 x \sqrt{-\hat{g}} \left( \frac{\hat{R}}{2\kappa^2} 
-\frac{1}{2} \hat{g}^{\mu\nu} \partial_{\mu} \phi \partial_{\nu} \phi 
- V(\phi) \right) \,,
\label{eq:ADD-FT-trace-1-a-III-B-02} \\ 
V(\phi) \Eqn{\equiv} \frac{\vartheta \hat{R} - F}{2 \kappa^2 \vartheta^2}\,,
\label{eq:ADD-FT-trace-1-a-III-B-03}
\end{eqnarray} 
where we have introduced a scalar field 
$\phi \equiv \left(\sqrt{3/2}/\kappa \right) \ln \vartheta$, and 
the hat means quantities in the Einstein frame. 
For $R^2$ inflation model with the 
action~(\ref{eq:ADD-FT-trace-1-a-III-B-01}), 
we find~\cite{CT-F-M}
\begin{equation}
V(\phi) = \frac{3 M_{\mathrm{s}}^2}{4 \kappa^2} 
\left(1-\exp \left(-\sqrt{2/3} \kappa \phi \right) \right)^2 \,.
\label{eq:ADD-FT-trace-1-a-III-B-04}
\end{equation}

We suppose 
the four-dimensional flat 
Friedmann-Lema\^{i}tre-Robertson-Walker (FLRW) 
space-time whose metric is given by 
\begin{equation}
d s^2 = dt^2 - a(t)^2 \sum_{i=1,2,3} \left(dx^i\right)^2\,. 
\label{eq:2.8}
\end{equation}
In this background, we choose the tetrad components as 
$e^a_\mu = \left(1,a\left(t\right),a\left(t\right),a\left(t\right)\right)$. 
Thus, we find $T=-6H^2$ with $H = \dot{a}/a$, where the dot denotes 
the time derivative. 

The values of quantities in terms of the curvature perturbations 
in $R^2$ inflation model are summarized 
as follows~\cite{Hinshaw:2012aka}. 
For $\phi \gg \phi_{\mathrm{f}}$ with $\phi(t=t_{\mathrm{f}}) \equiv \phi_{\mathrm{f}}$, where $t_{\mathrm{f}}$ is the end of inflation, 
slow-roll parameters $\epsilon$ and 
$\eta$ and the number of \textit{e}-folds $N_{\mathrm{a}}$ from 
the end of inflation at $t=t_{\mathrm{f}}$ to 
the era when the curvature perturbation with the comoving wave number 
$k=k_{\mathrm{WMAP}}$ crossed the horizon 
are described by~\cite{Hinshaw:2012aka, Salopek:1988qh}
%
\begin{equation}
\epsilon \equiv 
\frac{1}{2} \left( \frac{1}{V(\phi)} \frac{d V(\phi)}{d \phi} \right)^2 
= \frac{3}{4N_{\mathrm{a}}^2}\,,
\quad 
\eta \equiv 
\frac{1}{V(\phi)} \frac{d^2 V(\phi)}{d \phi^2} 
= - \frac{1}{N_{\mathrm{a}}}\,,
\quad 
N_{\mathrm{a}} \equiv
\int_{t}^{t_{\mathrm{f}}} H dt = 
\frac{3}{4} 
\exp \left( \sqrt{2/3} \kappa \phi \right)\,.
\label{eq:ADD-FT-trace-1-a-III-B-07}
\end{equation}
%
Furthermore, 
for the super-horizon modes $k \ll a H$, 
the scalar spectral index of primordial curvature perturbations 
$n_{\mathrm{s}}$ and the tensor-to-scalar ratio $r$ 
are expressed as~\cite{Mukhanov:1981xt, L-L}
%
\begin{equation}
n_{\mathrm{s}} -1 \equiv
\frac{d \ln \Delta_{\mathcal{R}}^2 (k)}{d \ln k} 
\biggm|_{k=k_{\mathrm{WMAP}}} 
= - 6 \epsilon + 2\eta \simeq -\frac{2}{N_{\mathrm{a}}}\,,
\quad \quad 
r = 
16 \epsilon = \frac{12}{N_{\mathrm{a}}^2} \,,
\label{eq:ADD-FT-trace-1-a-III-B-07}
\end{equation}
%
where $\Delta_{\mathcal{R}}^2$ is the amplitude of scalar modes of the primordial curvature perturbations at $k=0.002 \, \mathrm{Mpc}^{-1}$. 
The PLANCK results are $n_{\mathrm{s}} = 0.9603 \pm 0.0073$ (68\% errors) and 
$r < 0.11$ (95\% upper limit)~\cite{Ade:2013lta}. 
For instance, if $N_{\mathrm{a}}=50$, we find $n_{\mathrm{s}} = 0.96$ and 
$r=4.8 \times 10^{-3} \, (\ll 0.11)$. 

\subsection{Inflation in $T^2$ gravity} 

{}From the action in Eq.~(\ref{eq:ADD-FT-trace-1-a-III.1}), we see 
the reason why $T^2$ inflation does not work in $f(T)$ gravity. 
Our starting action with only gravity part and without matter is given by 
%
\begin{eqnarray} 
S \Eqn{=} \int d^4 x \abs{e} \frac{1}{2{\kappa}^2} \left( T+ \frac{1}{6M_{\mathrm{S}}^2} T^2 \right) \nonumber \\ 
\Eqn{=} \int d^4 x \abs{e} \frac{1}{2{\kappa}^2} \left[ \left(T+ 2\nabla^{\mu} T^{\rho}_{\verb| |\mu \rho}  \right) + \frac{1}{6M_{\mathrm{S}}^2} 
T^2 \right] \,,
\label{eq:ADD-FT-trace-1-a-III-C-01}
\end{eqnarray}
where in deriving the second equality, 
we have taken into account 
the fact that $\nabla^{\mu} T^{\rho}_{\verb| |\mu \rho}$ 
is a total derivative. 

In comparison of the action in Eq.~(\ref{eq:ADD-FT-trace-1-a13-IIIA-001}) 
for $R^2$ gravity with that in Eq.~(\ref{eq:ADD-FT-trace-1-a-III-C-01}) for 
$T^2$ gravity, 
we understand that the form of these actions are different 
from each other. Hence, it is expected that the cosmological evolutions 
in these two theories would be different. 
Indeed, for $R^2$ gravity, the de Sitter inflation cannot occur, 
whereas for $T^2$ gravity, the de Sitter inflation can be realized, 
as is investigated below. 

It is known that there does not exist the conformal transformation from 
an $f(T)$ gravity theory to a scalar field theory in pure teleparallelism, 
different from the case in $F(R)$ gravity~\cite{Yang:2010ji}. 
Thus, in $f(T)$ gravity, through the conformal transformation, 
it is impossible to move to the Einstein frame, i.e., 
the corresponding action consists of pure teleparallel term $T/\left(2\kappa^2 \right)$ plus the kinetic and potential terms of a scalar field, and examine the properties of inflation in the Einstein frame, 
as explained in the preceding subsection. 

To compare cosmological consequences for the action 
in Eq.~(\ref{eq:ADD-FT-trace-1-a-III-C-01}) in teleparallelism 
with those for an $R^2$ inflation model in convenient gravity, 
we examine solutions for the action 
in Eq.~(\ref{eq:ADD-FT-trace-1-a-III-C-01}) at the inflationary stage. 
In the flat FLRW background, 
the gravitational field equations can be written in the equivalent forms of 
those in general relativity: 
\be
\label{eq:4.1}
\frac{3}{\kappa^2} H^2 =  \rho_{\mathrm{eff}} \,, 
\quad - \frac{1}{\kappa^2} \left(2\dot{H} + 3 H^2 \right) 
= p_{\mathrm{eff}} \,, 
\ee 
where 
$\rho_{\mathrm{eff}}$ and $p_{\mathrm{eff}}$ are 
the effective energy density and pressure of the universe, respectively, 
given by 
\be
\label{eq:4.3}
\rho_{\mathrm{eff}} 
\equiv \frac{1}{2{\kappa}^2} \mathcal{I}\,, \quad 
p_{\mathrm{eff}} 
\equiv -\frac{1}{2{\kappa}^2} \left( 4\mathcal{J} + \mathcal{I} \right)\,,
\ee
with 
\begin{align}
\mathcal{I} \equiv& -T -f +2T f'
= 6 H^2 - f - 12 H^2 f' \,, 
\label{eq:IIB-Add-01} \\ 
\mathcal{J} \equiv& \left( 1 - f' -2T f'' \right) \dot{H}
= \left( 1 - f' + 12 H^2 f'' \right) \dot H \,. 
\label{eq:IIB-Add-02}
\end{align}
We should note that $\rho_{\mathrm{eff}}$ and $p_{\mathrm{eff}}$ 
in Eq.~(\ref{eq:4.3}) satisfy the standard continuity equation identically 
\begin{equation} 
\dot{\rho}_{\mathrm{eff}}+3H \left( 
\rho_{\mathrm{eff}} + p_{\mathrm{eff}} \right) = 0\,. 
\label{eq:4.5}
\end{equation} 
For the action 
in Eq.~(\ref{eq:ADD-FT-trace-1-a-III-C-01}), we have 
$f(T) = T+ \left[1/\left(6M_{\mathrm{S}}^2\right) \right] T^2$. 
It follows from Eq.~(\ref{eq:4.1}) that 
we have $0= f -2Tf'$, whose solutions are given by 
\begin{equation}  
H_{\mathrm{inf}} = 0\,, 
\quad 
\frac{M_{\mathrm{S}}}{\sqrt{3}} \, 
(= \mathrm{constant})\,. 
\label{eq:ADD-FT-trace-1-a3-IIIC-01}
\end{equation} 
Hence, we have a de Sitter solution at the inflationary stage, and 
eventually from the first equation in~(\ref{eq:4.1}) and 
Eq.~(\ref{eq:ADD-FT-trace-1-a3-IIIC-01}) 
we obtain 
\begin{equation} 
a(t) = a_{\mathrm{c}} \exp \left( H_{\mathrm{inf}} t \right) 
= a_{\mathrm{c}} \exp \left( \frac{M_{\mathrm{S}}}{\sqrt{3}} t \right) \,,
\label{eq:ADD-FT-trace-1-a3-IIIC-02}
\end{equation} 
where $a_{\mathrm{c}}$ is a constant. 
%
It is not known 
whether the de Sitter solution could be attractor or not, 
and therefore it is not clear 
if the inflation could be eternal or not is not. 
%
%
However, by including quantum effects 
such as the trace anomaly, 
the de Sitter solution is unstable against even small perturbations 
around this de Sitter solution, so that the de Sitter inflation 
can be over, as is seen in Sec.~IV B. 

On the other hand, it is known that for $F(R)$ gravity, at the de Sitter 
point, which is a solution for the case without matter, 
we have $F'(R) R - 2 F(R) =0$. If $F(R) = R+\left[ 1/\left(6M_{\mathrm{S}}^2 \right) \right] R^2$, we find $R = 0$. 
Moreover, we mention the case of an $R^2$ inflation model in the ordinary curvature gravity. In this case, the pure de Sitter inflation does not occur unlike in teleparallelism demonstrated above. 
Qualitatively, for $\left| \ddot{H}/\left(M_{\mathrm{S}}^2 H \right) \right| \ll 1$ and $\left| -\dot{H}^2/\left(M_{\mathrm{S}}^2 H^2 \right) \right| \ll 1$, we have $H = H_{\mathrm{i}} -\left(M_{\mathrm{S}}^2 /6 \right) \left(t-t_{\mathrm{i}} \right)$ and $a = a_{\mathrm{i}} \left[ H_{\mathrm{i}} \left(t-t_{\mathrm{i}} \right) -\left(M_{\mathrm{S}}^2 /12 \right) \left(t-t_{\mathrm{i}} \right)^2 \right]$, where $H_{\mathrm{i}}$ and $a_{\mathrm{i}}$ are the values of the Hubble parameter and scale factor at the initial time $t=t_{\mathrm{i}}$ of 
inflation, respectively. 
As a result, what is observed here is that 
the qualitative behavior of eternal inflation in $T^2$ gravity for teleparallelism is different from inflation in $R^2$ gravity.

\section{Trace anomaly}

In this section, we examine the cosmological influences of the trace anomaly 
on the energy density of the universe in modified gravity theories. 

\subsection{Effects of the trace anomaly} 

First of all, we should note that quantum effects of conformally-invariant fields on classical gravitational background are just the same as in $T$-gravity (i.e., pure teleparallelism) due to its equivalence to general relativity (GR). Furthermore, conformal invariance in teleparallelism is proved to be the same as in usual gravity. This requires non-minimal coupling with curvature~\cite{Bamba:2013jqa}. 

We now include the massless quantum effects by taking 
into account the trace anomaly $T_\mathrm{A}$, which has the following well-known form~\cite{Duff:1993wm}:
\be
\label{OVII}
T_\mathrm{A}= \tilde{b} \left(\mathcal{F} + \frac{2}{3}\Box R\right) + \tilde{b}' \mathcal{G} 
+ \tilde{b}'' \Box R\, ,
\ee
with $\tilde{b}$, $\tilde{b}'$, and $\tilde{b}''$ constants\footnote{Note that the prime in 
$\tilde{b}'$ and $\tilde{b}''$ is not the derivative with respect to $T$ but 
just a superscription as a notation.}, 
where $\mathcal{F}$ is the square of the 4D Weyl tensor, and 
$\mathcal{G}$ is the Gauss-Bonnet invariant, which are given by
\be
\label{GF}
\mathcal{F} = \frac{1}{3}R^2 -2 R_{\mu\nu}R^{\mu\nu}+
R_{\mu\nu\rho\sigma}R^{\mu\nu\rho\sigma}\, , \quad
\mathcal{G}=R^2 -4 R_{\mu\nu}R^{\mu\nu}+
R_{\mu\nu\rho\sigma}R^{\mu\nu\rho\sigma}\, .
\ee
When we consider $N$ scalars, $N_{1/2}$ spinors, $N_1$ vector fields, $N_2$ 
($=0$ or $1$) gravitons and $N_\mathrm{HD}$ higher-derivative conformal 
scalars, $\tilde{b}$ and $\tilde{b}'$ are given by 
\be
\label{bs}
\tilde{b} = \frac{N +6N_{1/2}+12N_1 + 611 N_2 - 8N_\mathrm{HD}}{120(4\pi)^2}
\, ,\quad
\tilde{b}' =- \frac{N+11N_{1/2}+62N_1 + 1411 N_2 -28 N_\mathrm{HD}}{360(4\pi)^2}\, .
\ee
We should note that $\tilde{b}>0$ and $\tilde{b}'<0$ for the usual matter, except the higher-derivative conformal scalars.
Notice that $\tilde{b}''$ can be shifted by the finite renormalization of the
local counterterm $R^2$, so $\tilde{b}''$ can be an arbitrary coefficient.
In the FLRW universe, we find
\be
\label{CA2}
R = 12 H^2 + 6 \dot H\, ,\quad 
\mathcal{F}=0\, ,\quad \mathcal{G}=24\left(H^4 + H^2 \dot H\right)\, .
\ee

Equations (\ref{eq:4.3}), (\ref{eq:IIB-Add-01}), and (\ref{eq:IIB-Add-02}) tell that the trace $T_\mathrm{T}$ of the energy-momentum tensor in the torsion 
scalar is given by 
\be
\label{FTt1}
T_\mathrm{T} = - \rho_\mathrm{T} + 3 p_\mathrm{T} 
= - \frac{2}{\kappa^2} \left( \mathcal{I} + 3\mathcal{J} \right) 
= - \frac{2}{\kappa^2} \left( 6H^2 + 3\dot H - f - 3\dot H f' - 12 H^2 f' 
+ 36 H^2 \dot H f'' \right) \, , 
\ee
where $\rho_{\mathrm{T}}$ and $p_{\mathrm{T}}$ are 
the energy density and pressure coming from the torsion scalar, respectively. 
Then by including the contribution from the trace anomaly, 
the equations in (\ref{eq:4.1}) are modified and we obtain
\be
\label{FTt2}
0 = - \frac{2}{\kappa^2} \left( - f - 3\dot H f' - 12 H^2 f' 
+ 36 H^2 \dot H f'' \right)
 - \left( \frac{2}{3} \tilde{b} + \tilde{b}'' \right) \left( \frac{d^2}{dt^2} + 3 H \frac{d}{dt} \right)
\left( 12 H^2 + 6 \dot H \right) + 24 \tilde{b}' \left( H^4 + H^2 \dot H \right)\, .
\ee
Especially if we consider the de Sitter space, where $H$ is a constant, 
$H=H_\mathrm{c}$, Eq.~(\ref{FTt2}) reduces to an algebraic equation:
\be
\label{FTt3}
0 = - \frac{2}{\kappa^2} \left( - f - 12 H_\mathrm{c}^2 f' \right)
+ 24 \tilde{b}' H_\mathrm{c}^4 \, .
\ee
When we do not include the contribution from the trace anomaly, that is, 
$\tilde{b}'=0$ in (\ref{FTt3}), 
\be
\label{FTt4} 
f_\mathrm{EH} = - 2 T + 12 H_\mathrm{c}^2\, ,
\ee
satisfies Eq.~(\ref{FTt3}) (note that $T=- 6 H_\mathrm{c}^2$), which is equivalent to 
the Einstein theory with a cosmological term. 
Particularly when 
\be
\label{FTt5}
f(T) = T + \beta T^n\, ,
\ee
where $\beta$ and $n$ are constants, 
{}from Eq.~(\ref{FTt3}) we find that the following relations are satisfied: 
\be
\label{FTt6}
H_\mathrm{c}^2 = 0\, , \quad 
\mathrm{and/or} \quad 
1+ \left(2n-1\right) \beta \left(-6H_\mathrm{c}^2 \right)^{n-1} 
+2H_\mathrm{inf}^2 \tilde{b}' \kappa^2 =0\, . 
\ee
For instance, when $n=2$, if the following relation is met: 
\be
\label{FTt7}
\frac{9 \beta}{\kappa^2} - \tilde{b}' > 0\, ,
\ee
there appears a solution corresponding to the de Sitter space, 
and the de Sitter (i.e., exponential) inflation can be realized. 
The condition~(\ref{FTt7}) is always satisfied as long as $\beta >0$, 
because $\tilde{b}' <0$ for the ordinary matter. 

On the other hand, in curvature gravity, 
the trace $T_\mathrm{C}$ of the energy-momentum tensor in the scalar curvature 
becomes 
$T_\mathrm{C} = 
\left( 1/\kappa^2 \right) 
\left[ -2\left( f'(R) R - f \right) +9\dot{f}'(R) H +12f'(R) H^2 
+3\ddot{f}'(R) + 6f'(R) \dot{H} - R \right]
$
with 
$R=6 \left( 2H^2 +\dot{H} \right)$, where the prime shows 
the derivative with respect to $R$, 
and 
the trace anomaly is expressed as 
$T_\mathrm{A} = -12 \tilde{b} \dot{H}^2 +24 \tilde{b}' \left(-\dot{H}^2 +H^2 \dot{H} + H^4 \right) -2\left(2\tilde{b} + 3\tilde{b}'' \right) \left(\dddot{H} +7H\ddot{H} + 4\dot{H}^2 + 
12H^2 \dot{H} \right)$~\cite{Nojiri:2005sx}. 
We take $F(R) = R + \tilde{\beta} R^{\tilde{n}}$ with $\tilde{\beta}$ and $\tilde{n}$ constants for the action in Eq.~(\ref{eq:ADD-FT-trace-1-a-III-B-01}). 
In this case, the de Sitter solution $H = \tilde{H}_\mathrm{c}$, 
where $\tilde{H}_\mathrm{c}$ is a constant, can be obtained if 
the following relation is satisfied: 
$0=\left(-\tilde{n} +2 \right) \tilde{\beta} \left(12 \tilde{H}_\mathrm{c}^2 \right)^{\tilde{n}}
+ 12 \tilde{H}_\mathrm{c}^2 + 24 \kappa^2 \tilde{b}' \tilde{H}_\mathrm{c}^4$. 
For $\tilde{n}=2$, this relation can be met only if $\tilde{H}_\mathrm{c}^2 = 0$ and/or 
$\tilde{b}' = -1/\left( 2\kappa^2 \tilde{H}_\mathrm{c}^2\right)$. 
In comparison with the case of teleparallelism, 
for curvature gravity, there exist only smaller regions of 
the model parameters to realize the de Sitter inflation.

\subsection{Instability of the de Sitter solution}

By following the procedure proposed in Ref.~\cite{alex}, 
we study the instability of the de Sitter solution leading to 
the end of inflation. 
For $f(T) = T+ \left[1/\left(6M_{\mathrm{S}}^2\right) \right] T^2$, 
the gravitational field equation (\ref{FTt2}) including the contribution 
of the trace anomaly reads 
\begin{eqnarray} 
0 \Eqn{=} \frac{6}{\kappa^2} 
\biggl[ 2H^2 -36 \frac{1}{\left(6M_{\mathrm{S}}^2\right)} H^4 + \dot{H} 
-36 \frac{1}{\left(6M_{\mathrm{S}}^2\right)} H^2 \dot{H} 
\nonumber \\
&&
{}- \left( \frac{2}{3} \tilde{b} + \tilde{b}'' \right) \kappa^2 
\left( \dddot{H} +7H \ddot{H} +12H^2 \dot{H} +4\dot{H}^2 \right) 
+4\tilde{b}' \kappa^2 
\left( H^4 + H^2 \dot{H} \right)
\biggr]\,.
\label{eq:add-FT-trace-1-a3-IVB-001}
\end{eqnarray}

We explore the small perturbation from the de Sitter solution as 
\begin{equation} 
H = H_\mathrm{inf} \left( 1 - \delta (t) \right)\,, 
\quad 
\delta \ll 1 \,,
\label{eq:add-FT-trace-1-a3-IVB-002} 
\end{equation} 
with $H_\mathrm{inf}$ the constant Hubble parameter at the inflationary 
stage for the case that the small perturbation vanishes ($\delta (t) =0$). 
Here, we only study the case that $\delta >0$, i.e., $H < H_\mathrm{inf}$ 
because for $\delta <0$, $H$ becomes larger in time and eventually diverges. 
By substituting this expression into 
Eq.~(\ref{eq:add-FT-trace-1-a3-IVB-001}) and taking only the leading (i.e., linear) order terms of $\delta$, we acquire 
\begin{eqnarray}
&&
\left( \frac{2}{3} \tilde{b} + \tilde{b}'' \right) \kappa^2 H_\mathrm{inf} \dddot{\delta} 
+7\left( \frac{2}{3} \tilde{b} + \tilde{b}'' \right) \kappa^2 H_\mathrm{inf}^2 \ddot{\delta} 
-\left[1-36 \left(\frac{1}{6M_{\mathrm{S}}^2}\right) H_\mathrm{inf}^2 
-12 \left( \frac{2}{3} \tilde{b} + \tilde{b}'' \right) \kappa^2 H_\mathrm{inf}^2 + 4\tilde{b}'\kappa^2 H_\mathrm{inf}^2 
\right]
H_\mathrm{inf} \dot{\delta} 
\nonumber \\
&&
{}+ 4 H_\mathrm{inf}^2 \left[ 1-36 \left(\frac{1}{6M_{\mathrm{S}}^2}\right) H_\mathrm{inf}^2 + 4 \tilde{b}' \kappa^2 H_\mathrm{inf}^2 \right] = 0\,.
\label{eq:add-FT-trace-1-a3-IVB-003}
\end{eqnarray}
We take the form of $\delta$ as $\delta = \exp \left(\chi t \right)$, where 
$\chi$ is a constant. If $\chi >0$, since the amplitude of $\delta$ 
grows as the cosmic time passes, the de Sitter solution 
becomes unstable, so that the universe can exit from the inflationary stage. 
By plugging the expression $\delta = \exp \left(\chi t \right)$ with 
Eq.~(\ref{eq:add-FT-trace-1-a3-IVB-003}), we have 
the cubic equation in terms of $\chi$
\begin{eqnarray}
&&
\chi^3 + 7H_\mathrm{inf} \chi^2 -\frac{
1 -36\left[1/\left(6M_{\mathrm{S}}^2\right)\right] H_\mathrm{inf}^2 
-12 \left[ \left(2/3\right) \tilde{b} + \tilde{b}'' \right] \kappa^2 H_\mathrm{inf}^2
+4\tilde{b}'\kappa^2 H_\mathrm{inf}^2
}{\left[ \left(2/3\right) \tilde{b} + \tilde{b}'' \right] \kappa^2} \chi 
\nonumber \\
&&
{}+\frac{4H_\mathrm{inf} \left\{ 1 -36\left[1/\left(6M_{\mathrm{S}}^2\right)\right] H_\mathrm{inf}^2 +4\tilde{b}'\kappa^2 H_\mathrm{inf}^2 \right\}}{\left[ \left(2/3\right) \tilde{b} + \tilde{b}'' \right] \kappa^2} = 0\,.
\label{eq:add-FT-trace-1-a3-IVB-004}
\end{eqnarray}
Provided that at the initial stage of inflation, $H$ varies 
slowly so that $\dddot{H} \ll H \ddot{H}$. In this case, by neglecting 
the $\chi^3$ term in Eq.~(\ref{eq:add-FT-trace-1-a3-IVB-004}) and 
solving it approximately, we find 
\begin{equation}   
\chi = \frac{1}{2} \left[ \left( 
\frac{7\mathcal{Q} -12 H_\mathrm{inf}^2}{7 H_\mathrm{inf}} \right)
 \pm \sqrt{\mathcal{D}} \right]\,, 
\label{eq:add-FT-trace-1-a3-IVB-005} 
\end{equation} 
where 
\begin{eqnarray}
\mathcal{Q} \Eqn{\equiv} 
\frac{1 -36\left[1/\left(6M_{\mathrm{S}}^2\right)\right] H_\mathrm{inf}^2 
+4\tilde{b}'\kappa^2 H_\mathrm{inf}^2}{7 \left[ \left(2/3\right) \tilde{b} + \tilde{b}'' \right] 
\kappa^2}\,,
\label{eq:add-FT-trace-1-a3-IVB-006} \\
\mathcal{D} \Eqn{\equiv} 
\left( \frac{7\mathcal{Q} -12 H_\mathrm{inf}^2}{7 H_\mathrm{inf}} \right)^2 
-16 \mathcal{Q}\,.
\label{eq:add-FT-trace-1-a3-IVB-007}
\end{eqnarray}
Since $\tilde{b}>0$, $\tilde{b}'<0$, and $\tilde{b}''$ is an arbitrary constant, $\mathcal{Q}$ can be negative. 
If $\mathcal{Q} <0$, we find $\mathcal{D} >0$, and 
$\chi$ with the `$+$' sign in Eq.~(\ref{eq:add-FT-trace-1-a3-IVB-005}) 
becomes positive. 
Consequently, 
when $\chi >0$, $\delta = \exp \left(\chi t \right)$ increases in time, 
and thus it follows from Eq.~(\ref{eq:add-FT-trace-1-a3-IVB-002}) that 
$H$ decreases, that is, the de Sitter solution becomes unstable. 

For $F(R) = R + \tilde{\beta} R^2$ (i.e., $\tilde{n} =2$), 
we examine the small perturbation around 
the de Sitter solution with Eq.~(\ref{eq:add-FT-trace-1-a3-IVB-002}). 
{}Substituting this into the gravitational field equations 
and only remaining the linear term in $\delta$, we obtain 
\begin{eqnarray}
&&
2\left[-18\tilde{\beta} 
+ \left(2 \tilde{b} + 3\tilde{b}'' \right) \kappa^2 \right] H_\mathrm{inf} 
\dddot{\delta} 
+2\left[-72\tilde{\beta} 
+7\left(2 \tilde{b} + 3\tilde{b}'' \right) \kappa^2 \right] H_\mathrm{inf}^2 
\ddot{\delta} 
\nonumber \\
&&
{}-6\left[ \left(1+72\tilde{\beta} H_\mathrm{inf}^2 \right) 
-4\left( 2 \tilde{b} -\tilde{b}' + 3\tilde{b}'' \right) \kappa^2 
H_\mathrm{inf}^2 
\right] H_\mathrm{inf} \dot{\delta} 
\nonumber \\
&&
{}-24 \left(1+4\tilde{b}' \kappa^2 H_\mathrm{inf}^2 \right) H_\mathrm{inf}^2 \delta 
+12\left(1+2\tilde{b}' \kappa^2 H_\mathrm{inf}^2 \right) H_\mathrm{inf}^2
= 0\,.
\label{eq:ADD-FT-trace-1-a5-IVB-01}
\end{eqnarray}
We consider the form $\delta = \exp \left(\tau t \right)$ with 
$\tau$ a constant. 
If $\tau >0$, the de Sitter solution becomes unstable, 
because the amplitude of $\delta$ increasing in time without the upper bound. 
It follows from the investigations in the last part of Sec.~IV A that for $\tilde{n}=2$ and $H_\mathrm{inf}^2 \neq 0$, if the de Sitter solution exists, 
we have $\tilde{b}' = -1/\left( 2\kappa^2 H_\mathrm{inf}^2\right)$. 
With this relation, we see that 
the first term in the third line of Eq.~(\ref{eq:ADD-FT-trace-1-a5-IVB-01}) is represented as $24 H_\mathrm{inf}^2 \delta$, and the last term in Eq.~(\ref{eq:ADD-FT-trace-1-a5-IVB-01}) is zero. 
By plugging $\delta = \exp \left(\tau t \right)$ with Eq.~(\ref{eq:ADD-FT-trace-1-a5-IVB-01}), we find 
\begin{eqnarray}
&&
2\left[-18\tilde{\beta} 
+ \left(2 \tilde{b} + 3\tilde{b}'' \right) \kappa^2 \right] H_\mathrm{inf} 
\tau^3 
+2\left[-72\tilde{\beta} 
+7\left(2 \tilde{b} + 3\tilde{b}'' \right) \kappa^2 \right] H_\mathrm{inf}^2 
\tau^2
\nonumber \\
&&
{}-6\left[ 72\tilde{\beta} H_\mathrm{inf}^2 
-4\left( 2 \tilde{b} + 3\tilde{b}'' \right) \kappa^2 
H_\mathrm{inf}^2 -1
\right] H_\mathrm{inf} \tau 
+24H_\mathrm{inf}^2 
= 0\,.
\label{eq:ADD-FT-trace-1-a5-IVB-02}
\end{eqnarray}
With the procedure used in teleparallelism shown above, 
we consider the case that the change of $H$ is slow enough for 
the $\tau^3$ term to be negligible in compared with 
the $\tau^2$ term. 
As a result, we acquire the approximate solution 
\begin{equation}   
\tau = \frac{3\mathcal{Z} 
\pm \sqrt{\tilde{\mathcal{D}}}}{2 \mathcal{W}}\,, 
\label{eq:ADD-FT-trace-1-a5-IVB-03} 
\end{equation} 
with
\begin{eqnarray} 
\mathcal{W} \Eqn{\equiv} 
-72\tilde{\beta} 
+7\left( 2 \tilde{b} + 3\tilde{b}'' \right) \kappa^2 \,,
\label{eq:ADD-FT-trace-1-a5-IVB-04} \\
\mathcal{Z} \Eqn{\equiv} 
\left[-\mathcal{W} + 3\left(2 \tilde{b} + 3\tilde{b}'' \right) \kappa^2 
\right] H_\mathrm{inf}^2 -1 \,,
\label{eq:ADD-FT-trace-1-a5-IVB-05} \\
\tilde{\mathcal{D}} \Eqn{\equiv} 
9\mathcal{Z}^2
-48 \mathcal{W} 
H_\mathrm{inf}^2 \,.
\label{eq:ADD-FT-trace-1-a5-IVB-06}
\end{eqnarray}
If we choose the `$-$' sign in Eq.~(\ref{eq:ADD-FT-trace-1-a5-IVB-03}), 
for $\mathcal{W} <0$ and $\mathcal{Z} >0$, we find $\tau >0$. 
When $\tau >0$, the amplitude of the perturbation diverges in time, 
and thus the de Sitter solution becomes unstable. 
This means the de Sitter inflation can end. 
Thus, we demonstrated that trace-anomaly driven inflation in 
$T^2$ or $R^2$ gravity has different behavior towards its exit.

\section{Trace-anomaly driven inflation in minimal massive bigravity} 

To check the universality of trace-anomaly driven inflation, we consider it in another version of modified gravity, so-called massive ghost-free bigravity, which has been very popular recently~\cite{deRham:2014zqa}. 

For the bimetric gravity, some cosmological solutions describing 
the accelerating universe have been investigated 
in several papers~\cite{Damour:2002wu, V-V, vonStrauss:2011mq, Akrami:2012vf, Berg:2012kn, NO-NOS-MMMNO, AKMS-LSS-BMMV-CMM}. 
Particularly, together with~Ref.~\cite{vonStrauss:2011mq}, 
the ghost-free, bigravity theory with cosmological solutions describing the accelerating universe has been examined in Ref.~\cite{Akrami:2012vf}. 
In the work, there has been presented an extensive and more detailed cosmological analysis of the model at the background level showing the existence of a late-time accelerated expansion without resorting to a cosmological constant. 
In this section, 
we include the trace anomaly to the bigravity models and show 
the existence of the solution describing de Sitter space-time, which may correspond to inflation in the early universe, so-called anomaly-driven inflation. 

We now investigate the minimal ghost-free bigravity model whose action is given by~\cite{DRG-DRGT, HR-HR}
\begin{align}
\label{bimetric2}
S_\mathrm{bi} =&M_g^2\int d^4x\sqrt{-\det g}\,R^{(g)}+M_f^2\int d^4x
\sqrt{-\det f}\,R^{(f)} \nonumber \\
&+2m^2 M_\mathrm{eff}^2 \int d^4x\sqrt{-\det g} \left( 3 - \tr \sqrt{g^{-1} f}
+ \det \sqrt{g^{-1} f} \right) + \int d^4 x \mathcal{L}_\mathrm{matter}
\left( g_{\mu\nu}, \Phi_i \right)\, .
\end{align}
Here 
$M_\mathrm{eff}$ is defined by
\be
\label{Meff}
\frac{1}{M_\mathrm{eff}^2} = \frac{1}{M_g^2} + \frac{1}{M_f^2}\, .
\ee
In (\ref{bimetric2}), $R^{(g)}$ and $R^{(f)}$ are the scalar curvatures for $g_{\mu \nu}$ and $f_{\mu \nu}$, respectively. 
The tensor $\sqrt{g^{-1} f}$ is defined by the square root of
$g^{\mu\rho} f_{\rho\nu}$, namely, 
$\left(\sqrt{g^{-1} f}\right)^\mu_{\ \rho} \left(\sqrt{g^{-1}
f}\right)^\rho_{\ \nu} = g^{\mu\rho} f_{\rho\nu}$. 

Then by the variation over $g_{\mu\nu}$, we get 
\begin{align}
\label{Fbi8}
0 =& M_g^2 \left( \frac{1}{2} g_{\mu\nu} R^{(g)} - R^{(g)}_{\mu\nu} \right)
+ m^2 M_\mathrm{eff}^2 \left\{ g_{\mu\nu} \left( 3 - \tr \sqrt{g^{-1} f}
\right)
+ \frac{1}{2} f_{\mu\rho} \left( \sqrt{ g^{-1} f } \right)^{-1\, \rho}_{\qquad 
\nu}
+ \frac{1}{2} f_{\nu\rho} \left( \sqrt{ g^{-1} f } \right)^{-1\, \rho}_{\qquad 
\mu} \right\} \nn
& + T_{\mathrm{matter}\, \mu\nu}\, .
\end{align}
On the other hand, through the variation over $f_{\mu\nu}$, we obtain 
\begin{align}
\label{Fbi9}
0 =& M_f^2 \left( \frac{1}{2} f_{\mu\nu} R^{(f)} - R^{(f)}_{\mu\nu} \right) \nn
& + m^2 M_\mathrm{eff}^2 \sqrt{ \det \left(f^{-1}g\right) } 
\left \{ - \frac{1}{2}f_{\mu\rho} \left( \sqrt{g^{-1} f} 
\right)^{\rho}_{\ \nu} - \frac{1}{2}f_{\nu\rho} \left( \sqrt{g^{-1} f} \right)^{\rho}_{\ \mu} 
+ \det \left( \sqrt{g^{-1} f} \right) f_{\mu\nu} \right\} \, .
\end{align}
By multiplying the covariant derivative $\nabla_g^\mu$ with respect to the 
metric $g$ by Eq.~(\ref{Fbi8}) and using the Bianchi identity 
$0=\nabla_g^\mu\left( \frac{1}{2} g_{\mu\nu} R^{(g)} - R^{(g)}_{\mu\nu} 
\right)$ and the conservation law $\nabla_g^\mu T_{\mathrm{matter}\, \mu\nu} =0$, 
we have 
\be
\label{identity1}
0 = - g_{\mu\nu} \nabla_g^\mu \left( \tr \sqrt{g^{-1} f} \right)
+ \frac{1}{2} \nabla_g^\mu \left\{ f_{\mu\rho} \left( \sqrt{ g^{-1} f } 
\right)^{-1\, \rho}_{\qquad \nu}
+ f_{\nu\rho} \left( \sqrt{ g^{-1} f } \right)^{-1\, \rho}_{\qquad \mu} 
\right\} \, .
\ee
Similarly by using the covariant derivative $\nabla_f^\mu$ with respect to the 
metric $f$, from (\ref{Fbi9}), we get 
\be
\label{identity2}
0 = \nabla_f^\mu \left[ \sqrt{ \det \left(f^{-1}g\right) } 
\left\{ - \frac{1}{2}\left( \sqrt{g^{-1} f} \right)^{ -1 \nu}_{\ \ \ \ \ \sigma} 
g^{\sigma\mu}  - \frac{1}{2}\left( \sqrt{g^{-1} f} \right)^{ -1 \mu}_{\ \ \ \ \sigma} 
g^{\sigma\nu} + \det \left( \sqrt{g^{-1} f} \right) f^{\mu\nu} \right\} \right]\, .
\ee
The identities (\ref{identity1}) and (\ref{identity2}) impose constraints 
on the solutions. 

We take the FLRW universes for the metric $g_{\mu\nu}$ 
and use the conformal time $t$. 
Furthermore, we suppose the form of the metric $f_{\mu\nu}$ as follows
\be
\label{Fbi10b}
ds_g^2 = \sum_{\mu,\nu=0}^3 g_{\mu\nu} dx^\mu dx^\nu
= a(t)^2 \left( - dt^2 + \sum_{i=1}^3 \left( dx^i \right)^2\right) \, ,\quad
ds_f^2 = \sum_{\mu,\nu=0}^3 f_{\mu\nu} dx^\mu dx^\nu
= - c(t)^2 dt^2 + b(t)^2 \sum_{i=1}^3 \left( dx^i \right)^2 \, .
\ee
We assume $a$, $b$, and $c$ are positive. 
In this case, from the $(t,t)$ component of (\ref{Fbi8}) we find
\be
\label{Fbi11b}
0 = - 3 M_g^2 H^2 - 3 m^2 M_\mathrm{eff}^2
\left( a^2 - ab \right) + a^2 \rho_\mathrm{matter} \, ,
\ee
and $(i,j)$ components yield
\be
\label{Fbi12b}
0 = M_g^2 \left( 2 \dot H + H^2 \right)
+  m^2 M_\mathrm{eff}^2 \left( 3a^2 - 2ab - ac \right)  
+ a^2 p_\mathrm{matter}\, .
\ee
Here $\rho_\mathrm{matter}$ and $p_\mathrm{matter}$ 
are the energy density and pressure of matter fields, respectively. 
On the other hand, the $(t,t)$ component of (\ref{Fbi9}) leads to 
\be
\label{Fbi13}
0 = - 3 M_f^2 K^2 +  m^2 M_\mathrm{eff}^2 c^2
\left ( 1 - \frac{a^3}{b^3} \right ) \, ,
\ee
and from $(i,j)$ components we find
\be
\label{Fbi14}
0 = M_f^2 \left( 2 \dot K + 3 K^2 - 2 LK \right)
+  m^2 M_\mathrm{eff}^2 \left( \frac{a^3c}{b^2} - c^2 \right)\, , 
\ee
with $K =\dot b / b$ and $L= \dot c / c$. 
Both Eqs.~(\ref{identity1}) and (\ref{identity2}) yield the identical 
equation:
\be
\label{identity3b}
cH = bK\ \mbox{or}\
\frac{c\dot a}{a} = \dot b\, .
\ee
If $\dot a \neq 0$, we obtain $c= a\dot b / \dot a$.
On the other hand, if $\dot a = 0$, we find $\dot b=0$, that is, $a$ and $b$ 
are constant and $c$ can be arbitrary. 

We 
suppose the de Sitter space-time for the metric $g_{\mu\nu}$ or $a$ 
and provided that 
$b$ and $c$ have the following forms: 
\be
\label{FbiTA1}
a = \frac{1}{H_0 t}\, ,\quad b= \frac{1}{K_0 t}\, , \quad c = \frac{1}{L_0 t}\, ,
\ee
with constants $H_0$, $K_0$, $L_0$. 
Then (\ref{identity3b}) gives 
\be
\label{FbiTA2}
K_0 = L_0\, ,
\ee
and therefore Eqs.~(\ref{Fbi11b}) and (\ref{Fbi12b}) 
have the following forms:
\begin{align}
\label{FbiTA3}
0 =& - 3 M_g^2 H_0^2 - 3 m^2 M_\mathrm{eff}^2
\left( 1 - \frac{L_0}{H_0} \right) + \rho_\mathrm{matter} \, , \\
\label{FbiTA4}
0 =& 3 M_g^2 H_0^2 
+  3 m^2 M_\mathrm{eff}^2 \left(  1 - \frac{L_0}{H_0} \right)  
+ p_\mathrm{matter}\, .
\end{align}
Equations (\ref{Fbi13}) and (\ref{Fbi14}) give an identical equation:
\be
\label{FbiTA5}
0 = - 3 M_f^2 L_0^2 +  m^2 M_\mathrm{eff}^2 
\left ( 1 - \frac{L_0^3}{H_0^3} \right ) \, .
\ee
Without matter, that is, when $\rho_\mathrm{matter}=p_\mathrm{matter}=0$, 
Eq.~(\ref{FbiTA3}) or Eq.~(\ref{FbiTA4}) conflicts with (\ref{FbiTA5}). 
In fact, if $H_0$ does not vanish, Eq.~(\ref{FbiTA3}) or Eq.~(\ref{FbiTA4}) 
with $\rho_\mathrm{matter}=p_\mathrm{matter}=0$ tells $L_0>H_0$, 
which conflicts with (\ref{FbiTA5}). 

We include the contribution from the trace anomaly in Eq.~(\ref{OVII})\footnote{Note that as the trace anomaly gives non-local and $R^2$-term contributions to the effective gravitational action, the ghost-free feature of massive bigravity may be lost.}. 
In the FLRW universe, we find
\be
\label{CA2}
R = \frac{6}{a^2} \left( H^2 + \dot H \right) \, ,\quad 
\mathcal{F}=0\, ,\quad \mathcal{G} = \frac{24 H^2 \dot H}{a^4} \, .
\ee
Then Eqs.~(\ref{FbiTA3}) and Eq.~(\ref{FbiTA4}) give 
\be
\label{FbiTA6}
0 = 12 M_g^2 H_0^2 + 12 m^2 M_\mathrm{eff}^2
\left( 1 - \frac{L_0}{H_0} \right) + 24{\tilde b}' H_0^4\, ,
\ee
which can be solved with respect to $L_0$: 
\be
\label{FbiTA7}
L_0 = H_0 + \frac{M_g^2 H_0^3}{m^2 M_\mathrm{eff}^2}
+ \frac{2 {\tilde b}' H_0^5}{m^2 M_\mathrm{eff}^2}\, .
\ee
Then substituting of $L_0$ in the above expression into 
(\ref{FbiTA5}), we obtain
\be
\label{FbiTA8}
0 = F\left( H_0^2 \right) \equiv - 3 M_f^2 H_0^2 \left( 1 + \frac{M_g^2 H_0^2}{m^2 M_\mathrm{eff}^2}
+ \frac{2 {\tilde b}' H_0^4}{m^2 M_\mathrm{eff}^2} \right)^2
+  m^2 M_\mathrm{eff}^2 
\left\{ 1 - \left( 1 + \frac{M_g^2 H_0^2}{m^2 M_\mathrm{eff}^2}
+ \frac{2 {\tilde b}' H_0^4}{m^2 M_\mathrm{eff}^2} \right)^3 \right\} \, .
\ee
If Eq.~(\ref{FbiTA8}) has a real and positive solution with respect to $H_0^2$ and $L_0$ in 
(\ref{FbiTA7}) is positive for the obtained $H_0$, the de Sitter space-time can be generated by the trace anomaly. 
When $H_0^2 \to 0$, $F\left( H_0^2 \right)$ behaves as 
$F\left( H_0^2 \right) \sim - 3 \left(M_f^2 + M_g^2 \right) H_0^2 <0$. 
On the  other hand, when $H_0^2 \to + \infty$, we find 
$F\left( H_0^2 \right) \sim -  \frac{8 \left({\tilde b}'\right)^3 H_0^{12}}{m^4 M_\mathrm{eff}^4}>0$. 
We should note ${\tilde b}'<0$ in general. 
Hence we find Eq.~(\ref{FbiTA8}) has surely a real and positive 
solution with respect to $H_0^2$. 
We also see that the value of $H_0^2$ gives a positive value of $L_0$ for (\ref{FbiTA7}). 
The function $ F\left( H_0^2 \right)$ can be rewritten as a function 
$\tilde F \left(L_0/H_0\right)$ of $L_0/H_0$ by using (\ref{FbiTA7}),
\be 
\label{FbiTA9}
\tilde F \left(L_0/H_0\right) = F\left( H_0^2 \right) 
=  - 3 M_f^2 H_0^2 \left( \frac{L_0}{H_0} \right)^2
+  m^2 M_\mathrm{eff}^2 
\left\{ 1 - \left( \frac{L_0}{H_0} \right)^3 \right\} \, .
\ee
Thus we find $\tilde F \left(0\right) = m^2 M_\mathrm{eff}^2> 0$ and 
when $L_0/H_0 \to + \infty$, $\tilde F \left(L_0/H_0\right) \sim  - m^2 M_\mathrm{eff}^2 
\left( \frac{L_0}{H_0} \right)^3 <0$. 
Therefore $\tilde F \left(L_0/H_0\right)$ vanishes for finite and positive $L_0/H_0$, 
which corresponds to $H_0^2$ satisfying $F\left( H_0^2 \right)=0$. 
Accordingly for $H_0^2$ which obeys $F\left( H_0^2 \right)=0$, $L_0$ is surely positive, 
and therefore the de Sitter space-time can be generated by the trace anomaly. 
We should note that the space-time described by $f_{\mu\nu}$ is also the de Sitter space whose length parameter is given by $1/L_0$. In (\ref{FbiTA9}), the first term is positive and hence the 
second term must be positive. This means $L_0<H_0$, that is, the expansion in 
the space-time described by $f_{\mu\nu}$ is slow, compared with the expansion in the space-time described by $g_{\mu\nu}$. 

In case of the standard Einstein gravity, instead of (\ref{FbiTA6}), we obtain
\be
\label{FbiTA10}
0 = 12 M_g^2 H_0^2 + 24{\tilde b}' H_0^4\, ,
\ee
whose solution is given by 
\be
\label{FbiTA11}
H_0^2 = - \frac{M_g^2}{2{\tilde b}'}\, .
\ee
In (\ref{FbiTA6}), the contribution from the mass term $12 m^2 M_\mathrm{eff}^2
\left( 1 - \frac{L_0}{H_0} \right)$ acts as a negative cosmological constant 
but the contribution makes $H_0^2$ or curvature larger than that in the Einstein gravity in (\ref{FbiTA11}). 

When the space-time is not the de Sitter space-time, Eq.~(\ref{FbiTA6}) is modified to be 
\be
\label{FbiTA12}
0 = 6 M_g^2 \left(H^2 + \dot H \right) + 12 m^2 M_\mathrm{eff}^2
\left( 12 a^2 - 9 ab - 3ac \right) + \frac{24{\tilde b}' H^2 \dot H}{a^2} 
 - \frac{2}{3}\frac{{\tilde b}''}{a^2} \frac{d}{dt} \left( a^2 \frac{d}{dt}\left(\frac{6}{a^2} 
\left( H^2 + \dot H \right) \right) \right) \, .
\ee
We now investigate if the de Sitter solution can be stable or not. 
Equation~(\ref{FbiTA12}) might be compared with Newton's equation of motion, 
$0= F - m \frac{d^2 x}{dt^2}$ in the classical mechanics. 
As a consequence, we can find the stability depends on the sign of ${\tilde b}''$ and the strength of the stability on the magnitude of ${\tilde b}''$~\cite{star, hawking1}. 
Even if the de Sitter solution is unstable, if we choose ${\tilde b}''$ large enough~\cite{star, alex}, which may correspond to the large mass in the classical mechanics, the de Sitter solution might be long-lived. Then, it represents realistic trace-anomaly driven inflation in massive bigravity. 

\section{Conclusions}

In the present Letter, we have studied trace-anomaly driven inflation in $T^2$ teleparallel gravity. In particular, it has clearly been demonstrated that 
the de Sitter inflation can be realized in $T^2$ gravity, while realistic quasi de Sitter inflation occurs in $R^2$ gravity. 

In addition, we have examined the effects of the trace anomaly on 
inflation in $T^2$ gravity. As a result, we have shown 
that the de Sitter inflation can occur and 
finally the universe can exit from it due to the de Sitter space instability 
solution coming from the trace anomaly. 
On the other hand, in convenient gravity, if the trace anomaly is taken into account, in $R^2$ gravity the realistic de Sitter inflation can happen. 
However, compared with the case in teleparallelism, 
the parameter regions leading to the de Sitter solution are more narrow. 
In this case, there are two sources for the exit from inflation: from the classical $R^2$-term and the similar term coming from the trace anomaly. 

Moreover, trace-anomaly driven inflation has been explored 
in minimal massive ghost-free bigravity. It has been observed that 
the de Sitter inflation can occur and continue for long enough time, 
and that eventually its instability 
induced by the trace anomaly can lead to the end of it. 
Thus, in this massive bigravity, trace-anomaly driven inflation 
with long enough duration can be realized. 
Finally, we remark that for massive bigravity, 
the contribution of the graviton mass plays a role of negative cosmological 
constant. 
It is interesting that one can use the universality of trace-anomaly driven inflation or its $R^2$-cousin for the unified description of inflation and late-time acceleration as it has been proposed in Ref.~\cite{Nojiri:2003ft}. Such a scenario may be realized in $F(R)$ bigravity~\cite{NO-NOS-MMMNO}.

\section*{Acknowledgments}

S.D.O. 
appreciates the Japan Society for the Promotion of Science (JSPS) 
Short Term Visitor Program S-13131 
and is grateful to the very warm hospitality at Nagoya University, 
where the work was initiated.
The work is supported in part
by the JSPS Grant-in-Aid for
Young Scientists (B) \# 25800136 (K.B.);\
that for Scientific Research
(S) \# 22224003 and (C) \# 23540296 (S.N.);\
and
MINECO (Spain), FIS2010-15640 and
AGAUR (Generalitat de Ca\-ta\-lu\-nya), contract 2009SGR-345,
and 
Min. Education and Science project (Russia) (S.D.O.). 


\end{document}